\documentclass[a4paper,11pt]{article}
\pdfoutput=1 
\usepackage{jheppub}
\usepackage{natbib}
\usepackage{dsfont}
\usepackage{braket}
\usepackage[T1]{fontenc} 
\title{\boldmath Holographic Interpretation of Relative State Complexity}

\author[a]{Alexander Yosifov,}
\author[b]{Aditya Iyer,}
\author[a]{and Lachezar Filipov}

\affiliation[a]{Space Research and Technology Institute, Bulgarian Academy of Sciences, Akad. G. Bonchev Street, Building 1, Sofia, 1113, Bulgaria}
\affiliation[b]{Department of Physics, University of Oxford, Parks Road, Oxford, OX1 3PU, UK}

\emailAdd{alexanderyyosifov@gmail.com}
\emailAdd{aditya.iyer@physics.ox.ac.uk}
\emailAdd{lfilipov@mail.space.bas.bg}

\abstract{We investigate a large-$N$ CFT in a high-energy pure state coupled to a small auxiliary system of $M$ weakly-interacting degrees of freedom, and argue the relative state complexity of the auxiliary system is holographically dual to an effective low-energy notion of computational cost in the bulk, \textit{i.e.} to the minimal depth of the quantum circuit required to decode its gravitational dual. In light of this, using Nielsen's approach, a new measure of quantum chaos in terms of the evolution of circuit complexity is proposed. It suggests that studying the evolution of circuit complexity of the auxiliary system can teach us about the chaotic properties of the large-$N$ CFT. This new diagnostic for quantum chaos has important implications for the interior dynamics of evaporating black holes as it implies the radiated Hawking cloud is pseudorandom.}

\begin{document} 
\maketitle
\flushbottom

\section{Introduction}

The AdS/CFT correspondence relates a $d$-dimensional boundary conformal field theory (CFT) to a bulk gravity theory in $(d+1)$-dimensional asymptotically Anti-de Sitter (AdS) spacetime. The existing bulk/boundary dictionary points to a deep relation between the concepts of quantum complexity, quantum information, chaos, and gravity -- all of which interplay in black holes.\\
Quantum complexity (herein referred to as \emph{complexity}) is usually defined in the literature as the minimum number of gates required to prepare a quantum state, assuming a universal gate set $\left\{ g_{i}\right\}$. Holographic complexity has been shown to admit two dualities, known as "complexity=volume" (CV-duality) and (ii) "complexity=action" (CA-duality). The CV-duality \cite{CV,CV2} dictates the evolution of complexity of a boundary CFT is gravitationally dual to the linear growth of the black hole interior,\footnote{We should note that, classically, the interior black hole volume continues to grow indefinitely, see Christodoulou-Rovelli \cite{rovelli}.} while the CA-duality \cite{CA, CA2} associates a family of weakly-coupled bulk degrees of freedom in the Wheeler-DeWitt (WDW) patch to the quantum state of the boundary CFT.\footnote{Using the classical action associated with the WDW patch in AdS spacetimes eliminates the ambiguities related to the size of the horizon relative to the AdS radius.}\\
Similarly important concept which we will focus on in this paper is \emph{relative state complexity}, roughly defined as the minimum number of gates required to prepare a target state $\left|\psi\right\rangle$ from a simpler reference state $\left|\psi_{0}\right\rangle$, assuming a universal gate set $\left\{ g_{i}\right\}$. Usually, for a two-sided AdS black hole the reference state is the thermofield-double (TFD) state, while for a one-sided AdS black hole it is the ground state. In fact, as we will assume throughout the paper for the case of a one-sided AdS black hole, the target state will simply be the time-evolved reference state $\ket{\psi}=U\ket{\psi_{0}}$, where $U$ denotes a unitary transformation.\footnote{Note that the entropic behavior of complexity \cite{SC} suggests the relative state complexity for any $\ket{\psi}$ and $\ket{\psi'}$ increases linearly in time and with the number of degrees of freedom.} Understandably, a lot of effort has been devoted \cite{MN,MN1,MN2} in the direction of investigating the evolution of complexity and relative state complexity from the perspective of geometry for which we have a richer toolbox.\\
Quantum chaos in strongly-coupled quantum systems of many degrees of freedom plays an important role in terms of information processing and strong thermalization which, on the other hand, can help us better understand quantum gravity. The semiclassical black hole structure, however, makes examining the interior highly non-trivial. Although different measures such as out-of-time-order correlators (OTOC) \cite{otoc1} and random matrix theory (RMT) \cite{rmt} exist, more recently, the incorporation of quantum information-theoretic tools in holography together with the established protocol for assigning computational costs to trajectories on the unitary manifold \cite{MN,MN1,MN2} have opened the possibility of probing the highly chaotic black hole interior from a new angle. Despite those advancements, however, we have a long way to go as we still do not fully understand what the CFT can teach us about the interior region. It would thus be useful to learn, for instance, how the fast scrambling of infalling matter is encoded in the CFT, and for that a more intuitive way of studying quantum chaos is needed. Given the widely adopted qubit description, the study of complexity and chaos in black holes has benefited tremendously lately from the (examined in Section \ref{Geometricalinterpretation}) geometric approach of Nielsen \textit{et al.} \cite{MN,MN1,MN2}.\\
Following the recent progress, in an attempt to extend the bulk/boundary dictionary and shed light onto the elusive nature of quantum chaos in strongly-coupled quantum systems, our goal in this paper is to propose, within AdS/CFT, a holographic interpretation of relative state complexity as a decoding task in the bulk, and demonstrate how it can be used as a novel measure of quantum chaos.\footnote{Similar proposals, relating quantum chaos and circuit complexity, have been recently explored \cite{magan, YF1, Kim, Masam}. Although there are technical differences in the approaches, all models agree that circuit complexity can be utilized as a new diagnostic of quantum chaos.} More precisely, employing Nielsen's complexity geometry framework, we investigate the evolution of relative state complexity of a small auxiliary system of $M$ weakly interacting degrees of freedom coupled to a large-$N$ CFT in a high-energy pure state, gravitationally dual to one-sided AdS black hole. We suggest the relative state complexity of the auxiliary system (i) holographically corresponds to the minimal depth of the quantum circuit necessary to decode its gravitational dual, and (ii) can furthermore be studied to derive the chaotic properties of the dual one-sided AdS black hole. In light of this, for the case of an evaporating black hole \cite{evap}, and in agreement with the conclusions reached by Kim-Tang-Preskill \cite{Preskill}, the quantum state of the emitted Hawking radiation is suggested to be pseudorandom, meaning an outside observer with limited computational resources (\textit{i.e.} using a polynomial-size quantum circuit) will not be able to decode it and thus distinguish it from a maximally mixed state. The supposed pseudorandomness of the Hawking radiation highlights the computational limitations of exterior observers in regard to decoding information processed by quantum chaotic systems and is reminiscent of the Harlow-Hayden computational cost argument \cite{HH1} in the context of decoding subfactors of the Hilbert space of the Hawking cloud in the firewall proposal.\\
Throughout the paper we will assume the idealized scenario in which the CFT evolution is described by a random quantum circuit\footnote{Random quantum circuits have been shown to be fast scramblers and thermalize quantum information in time logarithmic in the entropy \cite{random1,random2}.} acting on a large but finite $N$.\footnote{Throughout the paper we will work under the assumption that a black hole (holographically dual to a strongly-coupled large-$N$ CFT) is represented as a collection of $N$ qubits, where $S_{BH}=\frac{A}{4G_{N}\hbar}=N$.} Moreover, we will assume the random quantum circuit has a discrete time-step evolution $\Delta\tau$, dictated by a universal gate set of two-local gates.\footnote{For similar setup, see \cite{YF1}. Note that for a time interval $\Delta t \in [0,t]$, where $\Delta t \gg \Delta\tau$ we can divide $\Delta t$ into $j$ steps, where each is of order $\Delta\tau$.} 

\section{Complexity and Chaos}

In this Section we examine how complexity and chaos develop in large-$N$ random quantum circuits within AdS/CFT. We then put forward, employing Nielsen's approach, a geometrical interpretation of complexity and chaos, where the computational cost is given in terms of "distance" on a unitary manifold \cite{JM}.
Later, we demonstrate the same geometrical interpretation naturally
arises in the large-$N$ limit of the Sachdev-Ye-Kitaev (SYK) model.

\subsection{Complexity}

Complexity has quasi-periodic behavior. For a generic large-$N$ quantum system whose dynamics is dictated by a random quantum circuit, complexity is low for small $t$. Then, for $t\sim t_{*}$, due to the early chaotic dynamics, and bounded from above by the Lloyd's bound \eqref{eq:lloyd}, complexity grows exponentially, where for a system of $N$ degrees of freedom at temperature $T$

\begin{equation}
\label{scramblingcomplexity}
\mathcal{C}_{*}=N\log N
\end{equation}
Although the scrambling complexity is nowhere near the upper bound $\mathcal{C_*}\ll\mathcal{C}_{\text{max}}$, it is still substantial for $N\gg 1$. Having $e^{\lambda_{L}(t-t_{*})}\sim\mathcal{O}(1)$, where $\lambda_{L}$ is the Lyapunov exponent (\ref{eq:lyapunov}), at the scrambling time indicates the presence of chaos in the holographic CFT.\\
Later, for $t>t_{*}$, the exponential growth of complexity is saturated. After the scrambling time, it continues to increase but now linearly in the number of degrees of freedom \eqref{eq:lineargrowth}

\begin{equation}
\label{linearcomplexity}
\frac{d\mathcal{C}}{dt}\sim NT
\end{equation}
where both early- and late-time growth \eqref{scramblingcomplexity} and \eqref{linearcomplexity}, respectively, are restricted by the Lloyd's bound \cite{bound} \footnote{Being the fastest scramblers and the most chaotic objects in nature, black holes saturate the Lloyd's bound.}

\begin{equation}
\label{eq:lloyd}
\frac{d\mathcal{C}}{dt}\leq\frac{2M}{\pi\hbar}
\end{equation}
The liner growth continues for $t_{cr}\sim e^{N}$ (classical recurrence time) at which point the upper bound is reached

\begin{equation}
\mathcal{C}_{\text{max}}=\text{poly}(N)e^{N}    
\end{equation}
Complexity then remains at its maximum value for a quantum recurrence time $t_{qr}\sim e^{e^{N}}$ (doubly exponential in the entropy) and then begins to decrease.\\
The sharp transition in the dynamics at the scrambling time is well-motivated on both sides of the duality. In particular, if this exponential growth is saturated before the scrambling time, then this would indicate some yet unknown interior dynamics for AdS black holes which allows for faster information processing. On the other hand, exponential growth of complexity beyond the scrambling time would violate \eqref{eq:lloyd}, see Refs. \cite{JM,CE}.\\
Apparently, the scrambling time is of particular interest when studying large-$N$ chaotic systems. It is generally given as

\begin{equation}
\label{eq:scram}
t_{*}=\frac{\beta}{2\pi}\log N
\end{equation}
where $\beta\equiv T^{-1}$ is the inverse temperature. Having a two-local universal gate set, we can define the scrambling time as the time for a reduced density matrix to become approximately
thermal \cite{CA2}. Another definition, suitable for
strongly-coupled large-$N$ theories (dual to AdS black holes), is
the time for all degrees of freedom to indirectly interact. From the bulk/boundary equality of the Hilbert spaces, we see the
scrambling time is of particular importance for both AdS black holes, and high-temperature boundary CFTs since it indicates the presence of chaos.

\subsubsection{Complexity=Action}

In \cite{CA} Susskind \textit{et al.} suggested the complexity
of a CFT, living on the boundary of an asymptotically AdS spacetime, is dual to the action of a WDW patch in the bulk. The WDW patch is
defined as the union of the past and future light cones of a spacelike hypersurface, anchored at some boundary time. One can also think of it as the region spanned by the union of all spacelike hypersurfaces (\textit{i.e.} slices) anchored at some boundary time. In its most general form the CA-duality reads 

\begin{equation}
\label{eq:CA}
\mathcal{C}=\frac{I_{WDW}}{\pi\hbar}
\end{equation}
where the boundary complexity is dual to the action of the
\emph{entire} WDW region which extends deep within the AdS black hole
interior. Usually, the bulk $I_{WDW}$ contains an Einstein-Hilbert
action, and a York-Gibbons-Hawking boundary term.

The CA-duality is equally well-defined for both, one- and two-sided
AdS black holes. Suppose we have a CFT dual to a one-sided black hole
in the bulk, and pick an arbitrary boundary time $t$. The state of
the corresponding patch would be

\begin{equation}
\label{eq:CFTevol}
\left|\psi(t)\right\rangle =e^{-iHt}\left|\textrm{CFT}\right\rangle 
\end{equation}
where $H$ is a local Hamiltonian.\\
Eq. \eqref{eq:CFTevol} can be straightforwardly extended for the case of two entangled copies of a boundary CFT, dual to an eternal two-sided AdS black hole. In particular, picking $t_{L}$ and $t_{R}$ for the left and right AdS boundary, respectively, yields

\begin{equation}
\left|\psi(t_{L},t_{R})\right\rangle =e^{-i(H_{L}t_{L}+H_{R}t_{R})}\left|\textrm{TFD(\ensuremath{t_{L}},\ensuremath{t_{R}})}\right\rangle 
\end{equation}
where the TFD state is given as

\begin{equation}
\label{eq:TFD}
\left|\textrm{\textrm{TFD}(\ensuremath{t_{L}},\ensuremath{t_{R}})}\right\rangle \equiv\frac{1}{\sqrt{Z}}\sum_{n}e^{-\beta E_{n}/2}\left|n_{L}\right\rangle \otimes\left|n_{R}\right\rangle 
\end{equation}
here, $\left|n_{L,R}\right\rangle $ denotes the energy eigenstates,
and $\beta$ is the inverse temperature. 

Thus the conjectured CA-duality \eqref{eq:CA} suggests 
\begin{equation}
\mathcal{C}\left(\left|\psi(t_{L},t_{R})\right\rangle \right)=\frac{I_{WDW}}{\pi\hbar}
\end{equation}

\subsubsection{Complexity=Volume}

Initially proposed in \cite{CV} the CV-duality\footnote{Interestingly, the CA-duality can successfully reproduce the CV-duality without the need of adding any parameters by hand.} relates the complexity of a boundary CFT to the volume of a maximally-extended spacelike hypersurface behind the horizon 

\begin{equation}
\label{eq:CV1}
\mathcal{C}=\frac{V}{G_{N}l_{AdS}}
\end{equation}
where $l_{AdS}$ is the AdS radius.

Behind the horizon the volume of the hypersurface has been shown to
grow linearly like

\begin{equation}
V\sim ST
\end{equation}
where $S$ and $T$ are the entropy and the temperature of the black
hole, respectively.

As it was pointed out in \cite{LS,CV2}, however, the
CV-duality lacks the universality of the CA-duality since it requires
hand-put length scale. In particular, the relation \eqref{eq:CV1} is only valid
assuming the black hole is large compared to $l_{AdS}$. Otherwise,
for black holes smaller than $l_{AdS}$, (\ref{eq:CV1}) reads

\begin{equation}
\label{eq:CV}
\mathcal{C}=\frac{V}{G_{N}r_{+}}
\end{equation}
where $r_{+}$ is the horizon radius.

Note that $r_{+}$ depends on the mass of the black hole, and thus
has to be put ad hoc. Therefore, the CA-duality is considered more
universal, and can easily reproduce the CV relation.

\subsection{Chaos}

Chaos quantifies the sensitivity of a system to changes in the initial
conditions. The chaotic behavior of a strongly-coupled large-$N$
CFT manifests in the AdS bulk as fast scrambling \eqref{eq:scram}. A commonly used
way to probe chaos involves the use of out-of-time-order correlators, where for a strongly-coupled large-$N$ quantum
system at some fixed temperature $\beta$ 

\begin{equation}
\label{eq:OTOC}
\left\langle W(t)V(0)W(t)V(0)\right\rangle _{\beta}\thickapprox e^{\lambda_{L}t}
\end{equation}
Here, $\lambda_{L}$ is the Lyapunov exponent which is bounded from
above as 

\begin{equation}
\label{eq:lyapunov}
\lambda_{L}\leq\frac{2\pi}{\beta}
\end{equation}
$W$ and $V$ are simple Hermitian operators, where

\begin{equation}
W(t)\equiv e^{iHt}We^{-iHt}
\end{equation}
and $H$ is a local Hamiltonian.

At the scrambling time, due to chaos, the out-of-time-order correlator
\eqref{eq:OTOC}, up to a constant, decays exponentially \cite{CA2}

\begin{equation}
\left\langle W(t)V(0)W(t)V(0)\right\rangle _{\beta}=e^{\lambda_{L}(t-t_{*})}+\mathcal{O}(N^{-2})
\end{equation}
Thus, for $t>t_{*}$, regardless of $V$ and $W$, the correlator takes the following form

\begin{equation}
\left\langle VV\right\rangle \left\langle WW\right\rangle 
\end{equation}
This exponential decay is related to the rapid initial growth of the
commutator, which becomes highly non-trivial at the scrambling time.
For $t\ll t_{*}$ the commutator is suppressed by the large-$N$ term,
and it is both small, and approximately constant.\\
Notice that for chaotic quantum systems the behavior of the commutator
at early times is similar to that of complexity; namely, both quantities are initially low (and approximately constant for $t\ll t_{*}$), and later on undergo exponential growth, saturated at the scrambling time \cite{TA}.\footnote{The authors of \cite{TA} propose a new diagnostic for quantum chaos, where by tuning the interaction Hamiltonian, a timescale, associated with the transition between classical and chaotic dynamics, is derived.}\\
The growth of the commutator is illustrative of the growth of complexity of $W(t)$. As we later argue in the paper, one way to probe quantum chaos of a large-$N$ CFT would be to introduce a small auxiliary system, correlate it with the initial CFT, and examine the evolution of its relative state complexity.  

\section{Geometric Interpretation}
\label{Geometricalinterpretation}

\subsection{Complexity \& Chaos}

Following Nielsen \textit{et al.} \cite{MN,MN1,MN2}, we now put forward a geometrical interpretation of complexity and chaos to illustrate their relation to gravity. Here, complexity is interpreted as "distance" on a unitary manifold. In this language, "distance" essentially means computational cost. Note that due to the assumed bulk/boundary Hilbert space equality, distances have to be preserved across the duality. With that in mind, the geometric approach will later be used to study the gravitational dual of relative state complexity of a small weakly-interacting auxiliary quantum system entangled to a large-$N$ CFT.\\
Usually, the Fubini-Study metric \cite{fubinistudy} with its distance bound of

\begin{equation}
d\in\left[0,\pi/2\right]
\end{equation}
suffices when talking about quantum states orthogonality. One issue,
however, concerns the ease of saturating the bound. For the purpose
of studying complexity, the Fubini-Study metric cannot adequately describe its exponential upper bound, and thus has to be substituted.\\ 
That is why for studying complexity we employ a non-standard $2^{N}$-dimensional unitary manifold $U(2^{N})$. Here, a time-evolving quantum state defines a trajectory $s$ on the manifold, whose length naturally increases with time, corresponding to the state's increasing complexity. Similarly, for a pair of quantum states $\left|\psi\right\rangle$ and $\left|\psi'\right\rangle$, where $\left|\psi\right\rangle ,\left|\psi'\right\rangle \in U(2^{N})$,
each state defines its own trajectory on the manifold, where the distance between them linearly increases, and it is the geometrical analog of their increasing relative state complexity \cite{SC}. Moreover, the unitary manifold has intrinsic penalty factors which restrict quantum states from exploring more complex paths. Note that they are independent of the metric as a whole but rather depend on particular directions. Simple paths have $\mathcal{O}(1)$ penalties, while the more complex paths are exponentially suppressed by $e^{N}$ penalties. Obviously, the penalty factors are important because (i) they are related to the minimum possible complexity increase, associated with acting with a simple two-local gate (which geometrically can be interpreted as minimizing the geodesic length in the bulk), (ii) they define the local coordinates on the gravity side, and (iii) they define a natural notion of locality.\\
Similar to the SYK model examined below, a generic object of interest in this framework is the evolution operator

\begin{equation}
\label{eq:simple}
U(t)\equiv e^{-iHt}
\end{equation}
whose symmetry transforms as $U(s)=e^{-i\sum_{j}\theta_{j}(s)T_{j}}$, where $T_{j}$ is as defined in Section \ref{sec:syk}.\\
In a geometrical context, \eqref{eq:simple} formally reads

\begin{equation}
U(s)=\overleftarrow{\mathcal{P}}\exp\left(i\int^{s}ds H(s)\right)  
\end{equation}
where $\overleftarrow{\mathcal{P}}$ is a path-ordering operator and $H(s)$ is a local Hamiltonian which parameterizes a path $s$ on the unitary manifold

\begin{equation}
\label{eq:penalties}
H(s)=Y^{j}(s)\mathcal{G}_{i}
\end{equation}
where $Y^{j}(s)$ denotes the set of penalty factors which, when applied at every step (\textit{i.e.} point) along the geodesic, control its path, and $\mathcal{G}_{i}\equiv\left\{ g_{i}\right\}$ is a universal gate set. Proper choice of $\left\{Y^{j}\right\}$ ensures the computational task at hand is optimized. Geometrically, this translates to minimizing\footnote{It was demonstrated in \cite{VB} that the geodesic remains locally minimal for at least exponential time.} the local geodesic on $U(2^{N})$.\\
Working in a unitary manifold, the complexity of $U(t)$ is $e^{N}$, which is to say that $U(t)$ (here described as a point on $U(2^{N})$) can explore an $e^{N}$-dimensional state space. More precisely, we focus on the increase of complexity associated with applying $U(t)$ to an arbitrary pure quantum state $\left|\psi\right\rangle \in U(2^{N})$. Note that when we time-evolve a quantum state

\begin{equation}
\label{timeevolvingqstate}
e^{-iHt}\left|\psi\right\rangle 
\end{equation}
the corresponding complexity growth is independent of $\left|\psi\right\rangle$. Rather, the complexification is solely determined by the local Hamiltonian. Essentially, a time-evolving quantum state behaves like a non-relativistic particle moving across the unitary manifold. Where for a strongly-coupled large-$N$ system, due to the early chaotic dynamics for $t\sim t_{*}$ when the Lyapunov exponent is $\mathcal{O}(1)$, the distance traveled (\emph{i.e.} increase of complexity) is exponential

\begin{equation}
\label{expdistancegrowth}
d(t)=e^{2\lambda_{L}t_{*}}
\end{equation}
where at late times $t>t_{*}$ it saturates to an evolution linear in $N$

\begin{equation}
\label{eq:lineargrowth}
d(t)=Nt
\end{equation}
Evidently, the growth of the distance, traveled by a time-evolving
quantum state on the unitary manifold, is (i) exponential for $t\sim t_{*}$, (ii) linear in $N$ for $t>t_{*}$ until $t\sim t_{cr}$, and (iii) a function of the local Hamiltonian

\begin{equation}
\label{eq:complexityfunction}
\mathcal{C}\left(e^{-iHt}\left|\psi\right\rangle \right)\equiv\int F\left(H(t)\right)dt
\end{equation}
Given the discrete time-step evolution of the random quantum circuit, when describing the trajectory $s$, spanned by the quantum state on $U(2^{N})$, we need to specify how the local Hamiltonian acts at each time step $\Delta\tau$. For that purpose we define an instantaneous Hamiltonian $\widetilde{H}$, \textit{i.e.} a Hermitian operator which depicts the point-by-point evolution of the trajectory $s$

\begin{equation}
\label{eq:instHam}
\tilde H(s) = i \frac{dU(s)}{ds} U^\dagger(s)
\end{equation}
Here, an infinitesimal change along $s$ corresponds to a simple unitary operation, \textit{i.e.} acting with a two-local gate, and reads

\begin{equation}
\label{eq:infinitesimal}
U(s+ds)=e^{-i\tilde H(s)ds}U(s)    
\end{equation}
where geometrically, \eqref{eq:infinitesimal} can be expressed in the Schrodinger picture as, see Fig. \ref{fig:2}

\begin{equation}
\label{schrodinger}
i\frac{\ket{\phi(t_{1})}}{ds}=\tilde H(s)\ket{\phi(t_{0})}
\end{equation}
Despite being different than the classical Hamiltonian $H$, for the simplest case of an evolution operator $U(t)$ that we consider \eqref{eq:simple}, $\widetilde{H}=H$.\footnote{For a detailed description of the instantaneous Hamiltonian, see \cite{JM}.}\\
Therefore, a time-evolving quantum state, expressed in terms of the instantaneous Hamiltonian, reads

\begin{equation}
\label{eq:instH}
\tilde{H} \left( \rho(t) , \frac{d\rho}{dt} \right) = \sum_{j} \frac{\left[\rho(t),{\frac{d\rho}{dt}}\right]^j}{(j+1)!} \frac{d\rho(t)}{dt}
\end{equation}
where $\rho(t)$ is the density matrix corresponding to the time-evolving state and $j$ denotes the number of time steps each of interval $\sim\Delta\tau$. Essentially, \eqref{eq:instH} provides a microscopic step-by-step (\textit{i.e.} from $j$ to $j+1$) description of a time-evolving quantum state \eqref{timeevolvingqstate}, Fig. \ref{fig:2}. While geometrically, considering the intrinsic penalty factors of the unitary manifold \eqref{eq:penalties}, it corresponds to the "weighted choices" the geodesic makes each time step. That is, at each time step the geodesic evolution, dictated by the instantaneous Hamiltonian \eqref{eq:instHam}, \eqref{eq:infinitesimal}, \eqref{schrodinger}, and bounded by the penalty factors chooses a computationally economical (\textit{i.e.} mildly penalized) direction. Moreover, assuming the initial state $\ket{\phi(t_{0})}$ factorizes as

\begin{equation}
\label{factorizes}
\ket{\phi(t_{0})}=\underbrace{\ket{0}\otimes\dots\otimes\ket{0}}_{\text{j}}
\end{equation}
and given our choice of penalty factors \eqref{eq:penalties}, the instantaneous Hamiltonian can be expressed as \cite{magan}

\begin{equation}
\label{instafactorization}
\tilde{H}=\tilde{H}_{1}\otimes\mathds{1}\otimes\dots\otimes\mathds{1}+\mathds{1}\otimes\tilde{H}_{2}\otimes\mathds{1}\otimes\dots\otimes\mathds{1}+\dots+\mathds{1}\otimes\dots\otimes\mathds{1}\otimes\tilde{H}_{j}
\end{equation}
Therefore, following \eqref{eq:penalties} and \eqref{eq:instH}-\eqref{instafactorization}, the instantaneous Hamiltonian factorization yields

\begin{equation}
\braket{\phi(s)|\tilde{H}(s)|\phi(s)}=\sum_{j}\braket{\phi_{j}(s)|\tilde{H}_{j}(s)|\phi_{j}(s)}    
\end{equation}

\begin{figure}[tbp]
\centering
\includegraphics[scale=0.18]{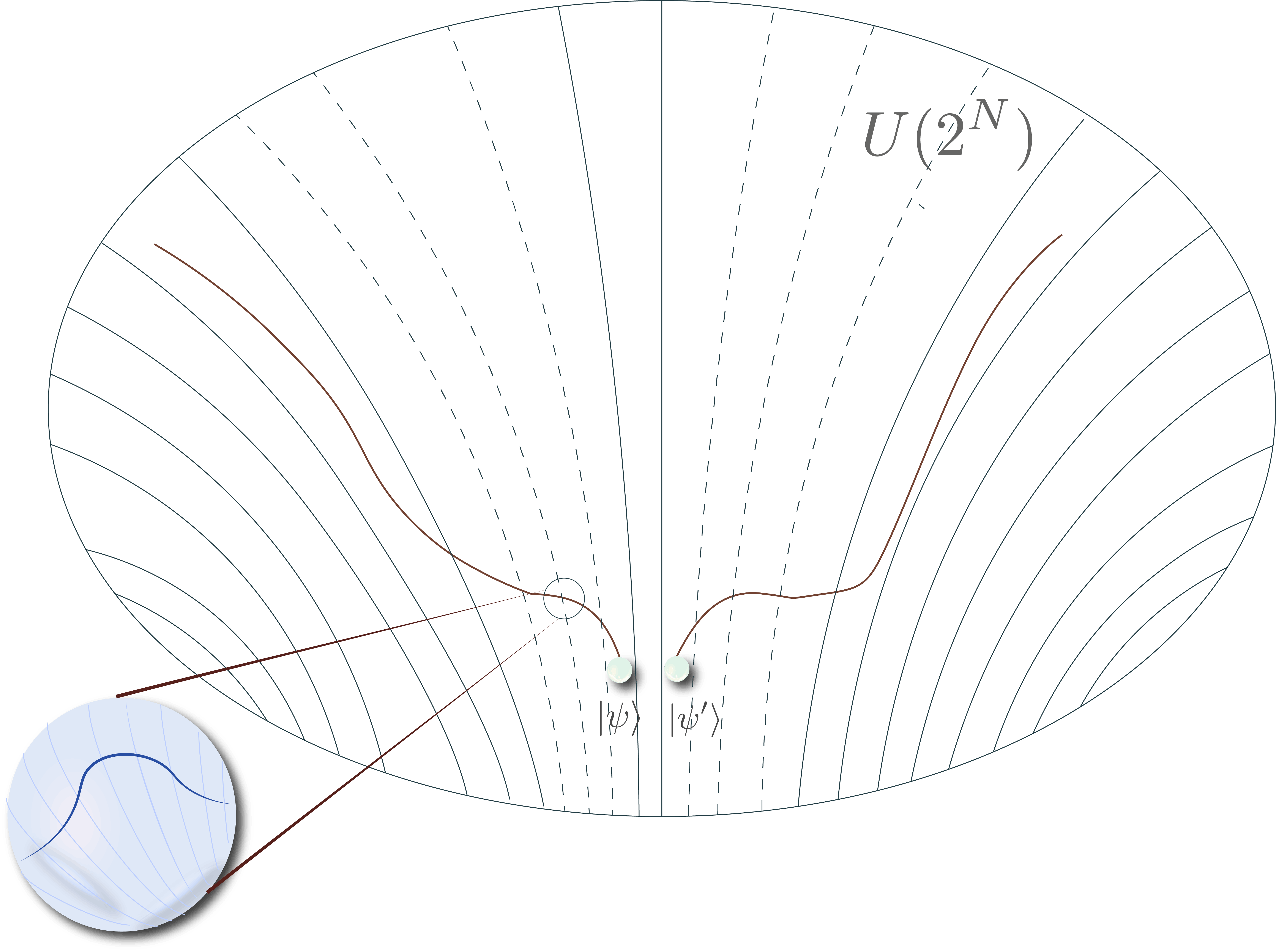}\caption{A pair of evolving quantum states on a unitary manifold which, geometrically, are described as geodesics. Initially, the quantum states are arbitrarily close, \emph{i.e.} low relative state complexity. Due to the chaotic early dynamics, however, their trajectories diverge. Later, for $t>t_{*}$ the distance between the states continues to grow but now linearly in $N$.}
\label{fig:1}
\end{figure}
In light of the proposed in Section \ref{section4} measure of quantum chaos in terms of circuit complexity, and to further illustrate the relation between complexity and geometry, suppose we
have a pair of arbitrarily close (\emph{i.e.} low relative state complexity) quantum states $\left|\psi\right\rangle$ and $\left|\psi'\right\rangle$, where $\left|\psi\right\rangle ,\left|\psi'\right\rangle \in U(2^{N})$, see Fig. \ref{fig:1}. Geometrically, the initial exponential growth of complexity \eqref{scramblingcomplexity} (indicative of chaos) can be interpreted as the rapid divergence of the trajectories of $\left|\psi\right\rangle$ and $\left|\psi'\right\rangle$. That is, for $t\sim t_{*}$, the number of simple steps $j$ between the states grows exponentially. As a result, assuming both states correspond to quantum systems of many degrees of freedom (as will be the case in Section \ref{section4}), the minimum size quantum circuit necessary to time-reverse the evolution of the system back to its initial low-complexity configuration grows immensely. Thus, the distance between the quantum states on $U(2^{N})$ increases exponentially \eqref{expdistancegrowth}, and as a result, the relative state complexity between the initially close states is highly non-trivial after only a scrambling time.\\
Even from this relatively straightforward setup, where an outside observer has access to all the relevant degrees of freedom, one can see that the computational cost of dealing with chaotic large-$N$ quantum systems, whose evolution is dictated by a random quantum circuit, grows quickly. Thus, extracting information from such complex systems for $t>t_{*}$ is very resource-demanding task. As we demonstrate in Section \ref{section4} for the more physically-relevant case of an AdS black hole entangled with its Hawking radiation in the context of quantifying quantum chaos in terms of circuit complexity, decoding the Hawking particles requires a superpolynomial-size quantum circuit. Meaning, the Hawking radiation is in a pseudorandom quantum state since it cannot be distinguished from a maximally-mixed state by any reasonable computation.\\
Before proceeding further, let us now briefly sketch an analog of the above argument in terms of particles in phase space $\mathcal{F}$, where similar conclusions regarding the evolution of relative state complexity at late times were derived \cite{magan,JM}. Here, nearby points $x(p,q)$ and $\Delta x(p,q)$ are depicted as quantum states

\begin{equation}
\begin{aligned}
&\ket{\psi}=\ket{p_{x}(t),q_{x}(t)}\\
&\ket{\psi'}=\ket{p_{\Delta x}(t),q_{\Delta x}(t)}
\end{aligned}    
\end{equation}
where in $\ket{\psi'}$ both the position $q$ and momentum $p$ are slightly perturbed.\\
In this context, a generic perturbation which can yield the evolution $\ket{\psi}\rightarrow\ket{\psi'}$, intuitively, reads

\begin{equation}
\label{eq:phasespaceunitary}
e^{i\Delta}=e^{i(p\Delta q-q\Delta p)}    
\end{equation}
where (\ref{eq:phasespaceunitary}) is a unitary operator which acts as a quantum circuit. Finding the complexity of $e^{i\Delta}$ can thus be used to calculate the relative state complexity between $\ket{\psi}$ and $\ket{\psi'}$. Geometrically, using a Hamiltonian functional similar to (\ref{eq:complexityfunction}), the complexity of $e^{i\Delta}$ can naturally be interpreted as the length of the minimum geodesic connecting the two points in $\mathcal{F}$. Notice, however, that for this to be true, the instantaneous Hamiltonian (\ref{eq:instHam}), giving the evolution of the Hamiltonian functional at each point along the geodesic, has to be considered for sufficiently small intervals.\\
Analogously to (\ref{timeevolvingqstate}) and (\ref{schrodinger}), time-evolving a quantum state is given as

\begin{equation}
\ket{\psi'}=e^{-iHt}e^{i\Delta}\ket{\psi}    
\end{equation}
where $\ket{\psi'}$ is perturbed, and its evolution can be expressed as

\begin{equation}
\ket{\psi'}=\ket{p(t)+\Delta p(t),q(t)+\Delta q(t)}    
\end{equation}
To summarize, evidently, similar to the above discussion, calculating relative state complexity between particles in $\mathcal{F}$ comes down to finding the minimum geodesic distance between their states in $\mathcal{F}$ (\ref{eq:infinitesimal}) which is achievable given (i) the Hamiltonian evolution is constant and (ii) the instantaneous Hamiltonian (\ref{eq:instHam}) is considered for sufficiently small time-intervals as to globally minimize the geodesic, \textit{i.e.} optimize the computational cost of acting with $e^{i\Delta}$.\\
Below we briefly show that the large-$N$ SYK model which is chaotic, holographically dual to $d$=2 quantum gravity, and describes the evolution of a chaotic Hamiltonian, admits identical late-time growth and geometrical interpretation.

\subsection{Sachdev-Ye-Kitaev}
\label{sec:syk}

Our goal is to further motivate the complexity and chaos evolution estimates and their geometrical interpretation. For this reason we proceed by very briefly looking at the geometrical approach to complexity in the standard large-$N$ SYK model describing a chaotic Hamiltonian in the Lie algebra formalism.\footnote{For a detailed take on this approach, see \cite{VB} and the references therein.} Consequently, we demonstrate the results in Section \eqref{Geometricalinterpretation} can easily be reproduced since the main objects have natural analogs within SYK.\\
Similar to the holographic case, to have a well-defined geometrical
approach to complexity in the SYK model, we need a basis which yields a notion of locality. We classify the Lie algebra of unitaries into two subgroups, \textit{i.e.} "easy" and "hard" directions on the unitary manifold. A natural notion of locality is introduced by the simple (\textit{i.e.} low complexity/computationally economical) generators $\left\{ T_{i}\right\}$ in the Lie algebra, where $\left\{ T_{i}\right\}=\gamma_{1}\gamma_{2}...\gamma_{n}$, which are analogous to quantum gates in the quantum information-theoretic approach. The Lie algebra generators are $k$-local, where for simplicity we set $k=2$, and strictly penalize the ($k>2$)-local ones. The $k=2$ restriction simply means no generator can act on more than two gamma matrices at a time. Here, the gamma matrices $\left\{ \gamma_{i}\right\} \sim N$ which satisfy the Clifford algebra, where $\left\{ \gamma_{\alpha},\gamma_{\beta}\right\} =2\delta_{\alpha\beta}$, play the role of qubits. Similar to the discussion above (see Section \eqref{Geometricalinterpretation}), locality implies the geodesic can only explore simple paths (\textit{i.e.} sourced by $k=2$ generators) on the unitary manifold, thus retaining its local minimum. Likewise, the restrictions on "easy" and "hard" directions are imposed by penalty factors which favor the use of $k=2$ generators. That way, within SYK, we have a straightforward notion of locality on the unitary manifold which assures the geodesic is locally minimized.\\
Therefore, considering \eqref{eq:penalties}, and assuming a universal two-local set of generators $\left\{T_{i}\right\}$, the path of a geodesic on $U(2^{N})$ is \cite{VB}

\begin{equation}
U(s)=\overleftarrow{\mathcal{P}}\exp\left(-i\int_{t_{1}}^{t_{2}}ds'V^{i}(s')T_{i}\right)
\end{equation}
where $V^{i}$ denotes the velocity terms. The unitary operator along
the trajectory can thus be given as

\begin{equation}
\frac{dU}{ds}=-iV^{i}(s)T_{i}U(s)
\end{equation}
Here, assuming a given direction is not strictly penalized, the velocity terms dictate the path of the geodesic. This implies that we need to consider the velocities for every value of $t$, \emph{i.e.} at each time step. Apparently, the velocities are equivalent to the instantaneous Hamiltonian in the holographic picture. In fact, we can make the relation even more precise and express the instantaneous Hamiltonian \eqref{eq:instHam} in terms
of the velocities in the tangent space as

\begin{equation}
\widetilde{H}(t)=\sum_{i}v^{i}T_{i}
\end{equation}
where only $\mathcal{O}(1)$ penalties are considered. This relation
ensures the unitary manifold is well-defined at each point along the
geodesic.\\
Evidently, the unitary evolution of the geodesic on the unitary manifold is sourced by (i) the velocities which control the path, and (ii) the two-local Lie algebra generators which ensure locality. The relation between geometry and complexity in the SYK model can thus be schematically expressed as \cite{Pol}

\begin{equation}
\mathcal{C}(e^{-iHt})=\int ds\left(G_{ij}V^{i}(s)V^{j}(s)\right)^{1/2}
\end{equation}
where $G_{ij}$ denotes the positive-definite bilinear metric form, and $H$ is the Hamiltonian.\\
Therefore, considering only the simple generators, the late-time growth of complexity reads \cite{VB, SYK}

\begin{equation}
\mathcal{C}(t)=\left(\sum_{n}(E_{n}t+2\pi k_{n})^{2}\right)^{1/2}
\end{equation}
Obviously, the behavior of complexity for a chaotic Hamiltonian in
the large-$N$ SYK model agrees with the holographic framework. Namely, at late times complexity increases linearly for time exponential in the number of degrees of freedom. Geometrically, a quantum state on the $2^{N}$-dimensional state space moves with velocity given by the sum over $E_{n}$, and its complexity is related to the distance traveled. 

\section{Relative State Complexity \& Pseudorandomness}
\label{section4}

\begin{figure}[tpb]
\label{figure2}
\centering
\includegraphics[scale=0.35]{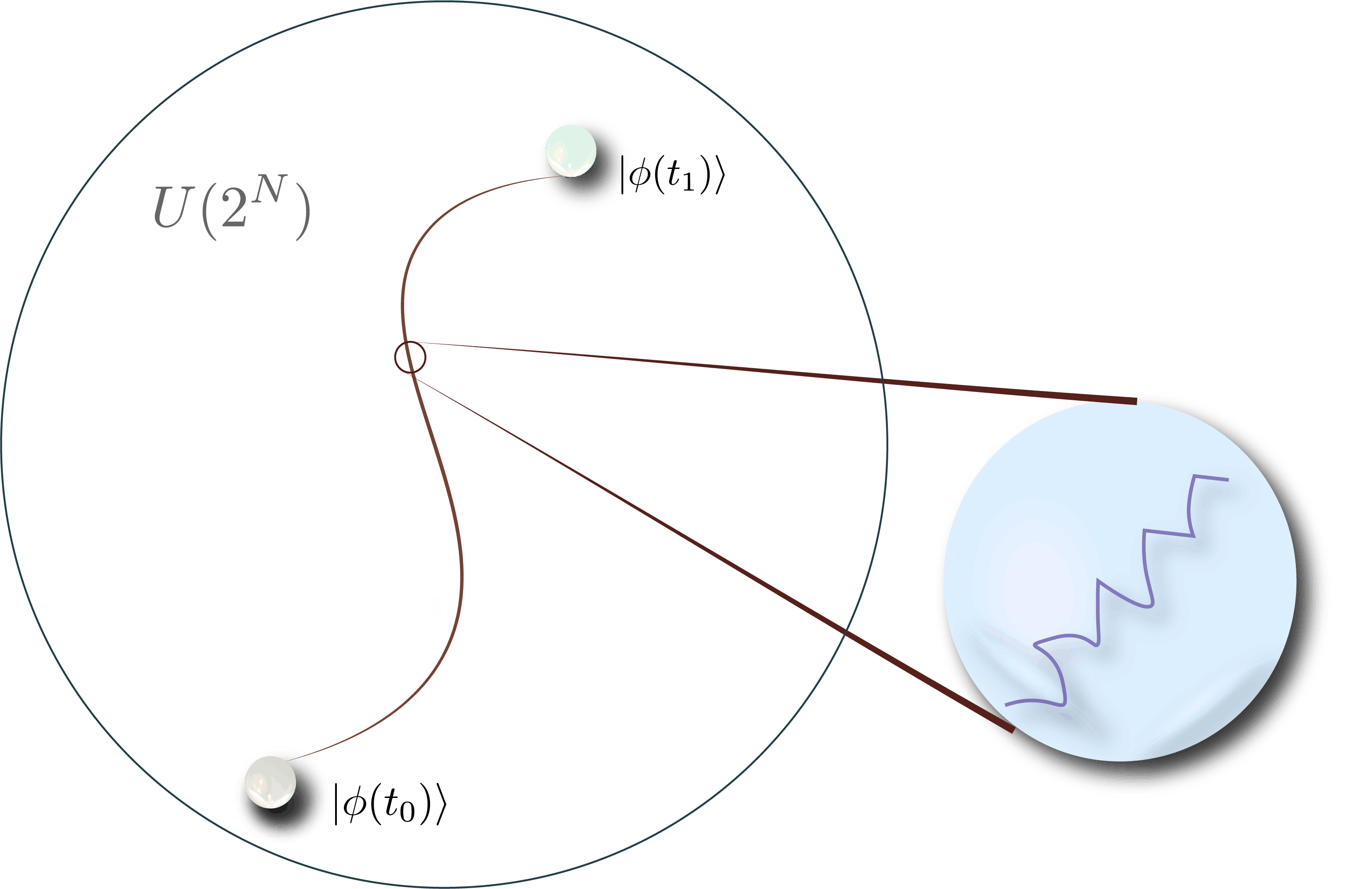}\caption{An evolving quantum state $\left|\phi\right\rangle$ from the identity $\ket{\phi(t_{0})}=\mathbb{I}=\mathds{1}$ to some later state $\left|\phi(t_{1})\right\rangle$ on a unitary manifold $U(2^{N})$. The line depicts the minimal geodesic between the quantum state at the two instances, $t_{0}$ and $t_{1}$. The distance between $t_{0}$ and $t_{1}$ corresponds to the relative state complexity, namely the computational cost associated with going from $\left|\phi(t_{1})\right\rangle$ to $\left|\phi(t_{0})\right\rangle$ \eqref{Uevol}. The zoomed-in region depicts the discrete evolution dictated at each step by the instantaneous Hamiltonian \eqref{eq:instH} and bounded by the penalty factors \eqref{eq:penalties}.}
\label{fig:2}
\end{figure}

In the current Section, using the geometric approach discussed in Section \ref{Geometricalinterpretation}, we examine a system of large-$N$ CFT in a high-energy pure state (dual to a one-sided AdS black hole) entangled to an auxiliary system of $M$ weekly-interacting degrees of freedom (dual to the Hawking radiation), where both systems are described as products of qubits with Hilbert spaces, respectively, $\mathcal{H}_{N}=2^{N}$ and $\mathcal{H}_{M}=2^{M}$, and we assume $N>M$.\footnote{It should be noted that the dual AdS black hole is "young," meaning it has evaporated much less than half of its initial degrees of freedom.} We study the evolution of complexity of the auxiliary system and argue its relative state complexity with respect to the identity is holographically dual to the minimum depth of the quantum circuit which can efficiently decode the Hawking radiation. We then suggest the growth of the relative state complexity can be utilized as a probe of the chaotic properties of the AdS black hole. Lastly, we argue the inability of any outside observer with reasonable computational resources to decode the Hawking cloud is indicative of its pseudorandom state.\\
It has been argued that ER=EPR \cite{EREPR}, meaning a pair
of CFTs in a nearly maximally-entangled state, living on the conformal AdS boundary, are dual to an eternal two-sided AdS black hole with smooth geometry behind the horizon. As is well known, AdS black holes with radius $\sim l_{AdS}$ do not evaporate due to the reflective conditions of the conformal boundary. However, as demonstrated by Raamsdonk \cite{evap}, weakly coupling a high-energy CFT to an auxiliary system perturbs the boundary conditions, and the dual AdS black hole evaporates, where following \cite{otoc1,rmt,random1,random2} the time scale associated with the beginning of the evaporation is set by the scrambling time.\footnote{The scrambling time is well-motivated even from a purely classical general relativity perspective as it is related to the relaxation time.} This way, black hole degrees of freedom leak to the adiabatically growing auxiliary system $M$ which now contains the radiation.\footnote{Note that if the auxiliary CFT is an exact copy of the original CFT, the two boundary theories are initially weakly correlated, and adiabatically evolve toward a TFD state \eqref{eq:TFD}.}\\
We here extend Raamsdonk's argument for the case of a large-$N$ high-energy CFT, entangled to a weakly-interacting
auxiliary system of $M$ degrees of freedom, and address a two-part question: \emph{What is the gravitational dual to the relative state complexity of the auxiliary system, and do the results have any implication to information loss/firewalls?}\\
As we show below, on the AdS boundary, the combined system of a large-$N$ CFT and a small-$M$ auxiliary CFT begins in a product state \eqref{eq:init}. By weakly correlating the two theories \eqref{interactionHamiltonian}, the initial product state adiabatically evolves to a TFD state \eqref{eq:TFD}. In the bulk, this is dual to a one-sided AdS black hole which at early times is in thermal equilibrium with its environment but then begins to evaporate, and at the Page time becomes maximally entangled with its Hawking radiation. We will now examine the relative state complexity of the auxiliary system with respect to the identity $\ket{\phi(t_{0})}=\mathbb{I}=\mathds{1}$, \emph{i.e.} relative to its value at $t=0$, where we will focus on the intermediate phase of the evolution while the black hole is still evaporating and the TFD state is not yet reached. Our claim is that this relative state complexity can be interpreted as being dual to an effective low-energy notion of computational cost in the bulk, namely to the minimal depth $\mathcal{D}_{\text{min}}$ of the quantum circuit, required to decode the Hawking quanta; the most efficient way to execute the computation \eqref{Uevol} in the form \eqref{eq:instH}. The depth of a quantum circuit gives the complexity-per-qubit measure of the computational task at hand, where in the random quantum circuit model we employ, the depth yields the number of time steps $j$ (or equivalently, the time $t$) needed to carry out \eqref{Uevol}. Thus, it effectively quantifies the relative state complexity of a computation. This measure is particularly useful when dealing with quantum systems composed of interacting qubits, especially in the complexity geometry approach.\footnote{Using the minimal depth of a quantum circuit as a measure of the computational cost associated with executing a task, \textit{e.g.} \eqref{Uevol}, resembles the Hartman-Maldacena tensor networks \cite{Hartman}. Tensor networks, like quantum circuits, have width and depth, and their evolution is very similar to that of circuits.}\\
More precisely, consider the following setup. On the
conformal AdS boundary we begin with a large-$N$ CFT in some high-energy pure state $\left|\psi_{0}\right\rangle$, and introduce a small-$M$ auxiliary system in its vacuum state $\left|0\right\rangle$, where $N>M$ and $N,M\gg 1$. The combined system is initially in a product state

\begin{equation}
\label{eq:init}
\left|\varPsi\right\rangle =\left|\psi_{0}\right\rangle^{\otimes N} \otimes\left|0\right\rangle^{\otimes M} 
\end{equation}
where $\left|\varPsi\right\rangle$ admits a Hilbert space factorization of the form

\begin{equation}
\mathcal{H}_{total}=\mathcal{H}_{N} \otimes \mathcal{H}_{M}    
\end{equation}
where $\mathcal{H}_{N}=\otimes_{i=1}^{N}\ket{\psi_{i}}$ and $\mathcal{H}_{M}=\otimes_{i=1}^{M}\ket{\phi_{i}}$. Suppose we now introduce an interaction Hamiltonian which entangles the two CFTs

\begin{equation}
\label{interactionHamiltonian}
H_{I}=\sum_{\gamma}C^{\gamma}_{NM}\mathcal{O}^{\gamma}_{N}\mathcal{O}^{\gamma}_{M}    
\end{equation}
where $\mathcal{O}^{\gamma}_{N}$ and $\mathcal{O}^{\gamma}_{M}$ are locally defined operators which only act on their respective Hilbert spaces, and $C^{\gamma}_{NM}$ denotes a family of coefficients.\\
The black hole interior $N$ is now purified by the exterior system $M$ (early and late Hawking radiation) and we assume, following AdS/CFT and being agnostic as to the exact mechanism as lies within the realm of quantum gravity, that $N$ is nonlocally encoded in $M$.\footnote{Effectively, the interaction Hamiltonian $H_{I}$ plays the role of an encoding map $H_{I}:\tilde{N}\rightarrow M$, where $\tilde{N}$ denotes an interior subregion which, assuming the validity of the equivalence principle, is purified by the late-time Hawking radiation.} That is, degrees of freedom are somehow transferred to the radiation reservoir $M$ \cite{harlow}. Intuitively, this would translate in the bulk to the one-sided AdS black hole starting to evaporate. Due to the transfer of modes, the auxiliary system is perturbed, and is no longer in the vacuum but instead in some \textit{typical} state $\left|\phi\right\rangle$, whose complexity we denote as $\mathcal{C}_{\phi}$. Moreover, the dimensionality of its Hilbert space $\mathcal{H}_{M}$ monotonically increases, corresponding to the steady growth of the number of degrees of freedom in $M$.\\
For small $t$, the complexity of the auxiliary CFT is low $\mathcal{C}_{\phi}\ll\mathcal{C}_{\text{max}}$, and due to the weak interactions of its degrees of freedom, grows linearly in $M$

\begin{equation}
\label{eq:linearC}
\frac{d\mathcal{C}_{\phi}}{dt}\sim MT
\end{equation}
where geometrically, \eqref{eq:linearC} implies the quantum state $\ket{\phi}$ of the auxiliary system $M$ moves in a particle-like manner across the unitary manifold with the length of the geodesic linearly increasing \eqref{eq:lineargrowth}, see Fig. \ref{fig:2}.\\
The evolution of the state $\ket{\phi}$ of the auxiliary system due to the highly non-trivial encoding map which non-locally relates the interior $N$ and exterior $M$ black hole regions, can schematically be given as $\ket{\phi(t_{1})}=U\ket{\phi(t_{0})}$, where $U$ is a unitary transformation by a polynomial-size random quantum circuit. That is, $\mathcal{C}_{\phi}$ quantifies how much more computationally demanding $\ket{\phi(t_{1})}$ is to decode; how many more steps $j$ it would take. For reasons we make precise below, we suggest this increase of $\mathcal{C}_{\phi}$ is reminiscent of the Harlow-Hayden firewall proposal \cite{HH1} and is equally unlikely to be computed.\footnote{Harlow-Hayden \cite{HH1} argued that AMPS' conjectured violation of the equivalence principle after Page time is computationally unrealizable for astrophysical black holes since it requires complicated measurements with superpolynomial-size quantum circuits on the emitted Hawking quanta the execution of which would take time exponential in the entropy. That is, no reasonable outside observer can verify that $\tilde{N}$ is purified by a subsystem $M$.} 

Unlike in \cite{HH1}, however, where Alice's goal was decoding only subfactors of the Hawking radiation in order to verify that early radiation is purified by modes in the black hole atmosphere, here the decoding task concerns applying unitary transformations to the auxiliary system to try to derive its initial state. Naively, calculating the relative state complexity may appear to be an easy task. Alice can simply apply $U^{\dagger}$ to the perturbed state $\left|\phi(t_{1})\right\rangle$ and time-reverse the operation 

\begin{equation}
\label{Uevol}
\left|\phi(t_{0})\right\rangle =U_{1}U_{2}U_{3}\cdots U_{j}\left|\phi(t_{1})\right\rangle 
\end{equation}
Notice, however, that such an operation -- unitary transformation by a polynomial-size random quantum circuit followed by a time-reversal operation, is conceptually similar to the quantum-mechanical measure of chaos, where a pair of identical states are evolved via slightly different Hamiltonians, resulting in exponential decay of their inner products \cite{chaos}. In our case, rather than changing the evolution operator, the slight difference comes from the encoding map. Namely, given the AdS black hole evaporates, in time $\Delta t$, where $t_{*}< \Delta t \ll S^{3/2}$, only a very small number $n$ of thermalized qubits will be transferred to $M$, where $n\ll M$. However, since black holes are notoriously good scramblers (fastest and most efficient in nature), and assuming Alice can only manipulate $M$, we argue those $n$ extra qubits will suffice to render the computation (\ref{Uevol}) unrealizable in time less than exponential in the black hole entropy. That is, no exterior observer with sub-superpolynomial size quantum circuit at their disposal can efficiently execute (\ref{Uevol}) and thus derive the initial state. Let's put the complex unitary encoding of $\tilde{N}$ into $M$ aside for a moment. Even under the assumption that $M$ is, to begin with, in a typical pure state $\rho_{M}=\ket{\psi}\bra{\psi}$ (chosen uniformly at random), it is still exponentially complicated for $\rho_{M}$ (considered with respect to a Haar distribution) to be distinguished from a maximally mixed state \cite{pseudo}. So to all computationally bounded observers the quantum state of $M$ will continue to appear maximally mixed. We can therefore conjecture that the computational cost of implementing \eqref{Uevol}; the relative state complexity of $\left|\phi(t_{1})\right\rangle$, can be interpreted as being gravitationally dual to \cite{CG}

\begin{equation}
\label{Dmin}
\mathcal{C}(\phi(t_1),\phi(t_0)) = \mathcal{D}_{\text{min}} \int_{t_0}^{t_1} F(\tilde{H}(s)) ds
\end{equation}
where $\tilde{H}(s)$ is as in \eqref{eq:instH} given some appropriate \eqref{eq:penalties}, and $\mathcal{D}_{\text{min}}$ denotes the minimum depth of the quantum circuit.\\
Since the auxiliary system $M$ plays the role of a Hawking cloud, we can apply similar analysis as in \cite{YF1,HH1} to try to estimate \eqref{Dmin} or at least put some constraints. Generally, the computational task Alice faces scales like $2^{M}$ for $m>0$, where $m$ denotes the leaking degrees of freedom to the auxiliary system. She could, of course, take different approaches to try to decode the auxiliary CFT, and hence execute \eqref{Uevol} efficiently. For instance, we imagine Alice could manipulate the degrees of freedom of the adiabatically growing auxiliary system, and engineer them into individual sets. She could then apply, in succession or in parallel, unitary transformation to the different sets in any arbitrary order she wishes. As it was demonstrated in \cite{HH1}, however, this procedure of limiting the unitary transformation to any particular group of degrees of freedom is especially complicated. Even more so, given $m>0$, meaning there is a non-local map encoding $\tilde{N}$ into $M$, multiple such limiting transformations have to be considered, further complicating the computation. Another decoding approach Alice could take is to make specific gates act on particular groups of degrees of freedom. Or similarly, connect different groups to specific gate subsets, and choose which groups to be acted on and when. Establishing any such connections would obviously require introducing extra degrees of freedom which scale as $e^{M}$.\\
Evidently, even assuming $\tilde{N}$ is encoded in $M$, meaning there is a unitary non-local transfer of information to the exterior region, because black holes are such efficient scramblers, an outside observer with polynomial-size quantum circuit and access only to $M$ will not be able to read it. That is to say that $\tilde{N}$ is hidden to computationally bounded exterior observers.\footnote{We should note, however, that if Alice had access to all the relevant degrees of freedom, \textit{e.g.} she waits for the black hole to evaporate completely, she would be able to efficiently decode $M$.} As far as Alice is concerned, performing a quantum computation on $M$, she will not be able to distinguish $\tilde{N}$ from the maximally mixed state of the thermal bath of radiation $M$. This implies that for ordinary observers the execution of (\ref{Dmin}), \textit{i.e.} calculating the relative state complexity, is computationally unrealizable. Even if we suppose that an $n$-qubit pure state, where $n\ll N$ is encoded in $M$, for any reasonable quantum system $N$ which is also a fast scrambler, the probability of Alice distinguishing the $n$ qubits from the maximally mixed radiation with precision (error tolerance) logarithmic in the number of gates of her quantum circuit is $\sim e^{-2^{n}}$. Thus, for generic $M\gg 1$, decoding it and reading the extra $n$ qubits is exponentially unlikely. Therefore, this rapid growth of the computational complexity of executing (\ref{Dmin}) is indicative of the chaotic  black hole dynamics. Said otherwise, quantum computation seems to very well protect the interior spacetime. This robustness of the semiclassical spacetime is usually discussed in the following context. Suppose we have a pair of entangled black holes in the TFD state (\ref{eq:TFD}). Alice is outside of her black hole while Bob has already crossed the horizon on his side. For Alice to perturb her CFT and send a high-energy Planckian messages to Bob (create a firewall behind his horizon) would require her to either act with an exponentially complex quantum circuit or apply a highly fine-tuned future precursor operator, thus making the computation unrealizable for astrophysical black holes \cite{message,message2}.\\
The current work may be considered as an extension of \cite{YF1} where we initially studied this new measure of chaos in strongly-coupled quantum systems of many degrees of freedom in terms of circuit complexity. We demonstrated that, due to the chaotic dynamics of the black hole and its causal semiclassical structure, Alice cannot decode the Hawking subsystem faster than time exponential in the entropy. Furthermore, we showed that Alice has two options, she can either act with a maximally complex unitary operator or act with future precursor operators to the perturbed state, and rely on extreme fine-tuning, where both options were argued to be computationally unreasonable for astrophysical black holes formed by sudden collapse. The exponential growth at the scrambling time of the minimum number of time steps $j$ (as defined in \eqref{eq:instH}) required to implement \eqref{Uevol}, and thus calculate the relative state complexity, indicates the presence of chaos in $N$. We showed that by studying the circuit complexity, we can learn about the efficiency of the information processing, and the chaotic dynamics of the black hole interior. Furthermore, by using the circuit complexity as a measure of quantum chaos, we demonstrated that the Hawking radiation is pseudorandom. Namely, assuming there are information-carrying particles among the radiated thermal Hawking quanta, they are scrambled beyond recognition given Alice does not have superpolynomial computational resources, and can only act on $M$. We have thus made the case that the $2^{k+m+r}$ bound proposed by Harlow-Hayden \cite{HH1} holds strong even for young black holes.\footnote{The results we derived in \cite{YF1} have since been supported in \cite{TA}, and more recently in \cite{Preskill}. Moreover, in terms of the AMPS' paradox, our work reaffirms the expectations that creating a firewall by an outside observer is notoriously difficult and computationally-demanding task.}\\

\section{Conclusions}

In summary, we examined the case of a small auxiliary system of $M$ degrees of freedom weakly coupled to a large-$N$ high-temperature CFT, whose bulk dual is an evaporating one-sided AdS black hole entangled to the Hawking cloud. We demonstrated that the natural linear increase of the relative state complexity of the auxiliary CFT, \eqref{eq:linearC}, with respect to the identity, \emph{i.e.} between $\left|\phi(t_{0})\right\rangle$ and $\left|\phi(t_{1})\right\rangle$ could be interpreted as being dual to a low-energy notion of computational cost (decoding task) in the bulk. That is, to the minimal depth of the quantum circuit required to decode the auxiliary CFT, namely to execute \eqref{Uevol} in the form \eqref{eq:instH}. In particular, the auxiliary system, playing the role of a Hawking cloud in the bulk, gets harder to decode with time, corresponding to the increasing relative state complexity of its boundary dual. We showed that the inability of any computationally bounded exterior observer to decode the Hawking radiation demonstrates its pseudorandom state, and can further be used as a measure of quantum chaos in terms of circuit complexity.  

\section*{Acknowledgement}
AY is thankful to the University of Oxford for the hospitality. We would like to acknowledge, in particular, the numerous insightful discussions with Vlatko Vedral. The authors are also grateful to Jinzhao Sun for his help with the figures.

\end{document}